# Heterodiffusion coefficients in $\alpha$-iron

Vassiliki Katsika-Tsigourakou[*] and Efthimios S. Skordas

Department of Solid State Physics, Faculty of Physics, University of Athens,
Panepistimiopolis, 157 84 Zografos, Greece

**Abstract**

The diffusion of tungsten in $\alpha$-iron is important for the application of ferritic-iron alloys to thermal power plants. These data, over a wide temperature range across the Curie temperature, have been recently reported. We show that these diffusion coefficients can be satisfactory reproduced in terms of the bulk elastic and expansivity data by means of a thermodynamical model that interconnects point defects parameters with bulk qualities.

---

[*] Email: vkatsik@phys.uoa.gr

## 1. Introduction

In a recent review [1], the models that interconnect point defect parameters with bulk properties have been presented. Chief among these, the so called $cB\Omega$ model [2-7], which suggests that the defect Gibbs energy $g^i$ (where $i$ denotes the corresponding process, i.e., defect formation, f, migration, m, or self-diffusion activation, act) is proportional to the isothermal bulk modulus $B$ and the mean volume $\Omega$ per atom. After investigating a large variety of solids, it was finally concluded [1] that the $cB\Omega$ model leads to results that are in agreement with the experimental data.

The above review [1] was crossed with the publication by Takemoto et al. [8] of tracer diffusion coefficients of $^{181}$W in $\alpha$-iron in the temperature range between 833 and 1173 K using serial sputter-microsectioning method. These data in high purity $\alpha$-iron, over a wide temperature range across the Curie temperature (T$_C$=1043 K), were considered to be important primarily in two respects: First, it is well known that a small addition of large-size elements, such as W, Nb and Mo, into iron increases considerably creep strength of the steels. Second, earlier measurements were limited to temperatures above 973 K, while data at lower temperatures, in particular around 900 K are necessary for the application of ferritic-iron alloys to thermal power plants. It is the object of this paper to investigate whether the $cB\Omega$ model can reproduce these important diffusion data.

We clarify that, as already mentioned in Ref. [1], the aforementioned "elastic" models (in the sense that $g^i$ is interrelated with bulk elastic data) have recently attracted the interest in view of the following facts: A challenging suggestion has been forwarded [9, 10] that these "elastic" models may provide a basis for the understanding of the non-Arrhenius temperature dependence of the viscosity of the glass forming liquids when the glass transition [11] is approached. Furthermore, it was found [12] that, in a certain class of high $T_C$-superconductors, the formation energy for Schottky defects follows the expectations of the $cB\Omega$ model. Finally we note that, when applying uniaxial stress in ionic crystals electric signals are produced which have parameters that are consistent with the $cB\Omega$ model [13]. This is important for understanding the generation of precursory electric signals that are measured before seismic events [14-16].

## 2. The diffusion coefficients

The diffusion coefficient D, if a single diffusion mechanism is operating in mono-atomic crystals, is described in terms of the activation Gibbs energy $g^{act}$, as [7]:

$$D = f\alpha^2 \nu \exp(-\frac{g^{act}}{k_B T}) \qquad (1)$$

where $f$ is a numerical constant depending on the diffusion mechanism and the structure, $\alpha$ stands for the lattice constant, $\nu$ the attempt frequency which for the self-diffusion activation process is of the order of the Debye frequency $\nu_D$ and $k_B$ the usual Boltzmann constant.

The activation entropy $s^{act}$ and the activation enthalpy $h^{act}$ are defined [7] in terms of $g^{act}$ as follows:

$$s^{act} = -\frac{dg^{act}}{dT}\bigg|_P \qquad (2)$$

$$h^{act} = g^{act} - T\frac{dg^{act}}{dT}\bigg|_P, \text{, and hence } h^{act} = g^{act} + Ts^{act} \qquad (3)$$

If the plot $\ell n D$ versus $1/T$ is linear, both $h^{act}$ and $s^{act}$ are temperature independent and then Eq.(1) can be written as:

$$D = D_0 \exp(-\frac{h^{act}}{k_B T}) \qquad (4)$$

where $D_0$ is given by

$$D_0 = f\alpha^2 \nu \exp(\frac{s^{act}}{k_B}) \qquad (5)$$

Let us now write $D$ in terms of the $cB\Omega$ model. Since the defect Gibbs energy $g^i$ is interconnected with the bulk properties of the solid through the relation:

$$g^{act} = c^{act} B\Omega \qquad (6)$$

where $c^{act}$ is a dimensionless constant, by substituting Eq. (6) into equation (1) we get

$$D = f\alpha^2 \nu \exp(-\frac{c^{act} B\Omega}{k_B T}) \qquad (7)$$

This relation, enables the calculation of $D$ at any temperature provided that elastic and expansivity data are available and that $c^{act}$ has been determined from a single measurement (i.e., once the value $D_1$ has been found experimentally at a temperature $T_1$, the value of $c^{act}$ can be determined since the pre-exponential factor $f\alpha^2\nu$ is approximately known [17] because $\nu$ can be roughly estimated as it will be explained below). The values of $s^{act}$ and $h^{act}$ can then be directly calculated at any temperature by means of the following equations that result upon inserting Eq.(6) into Eq.(2) and (3):

$$s^{act} = -c^i \Omega (\beta B + \frac{dB}{dT}\bigg|_P) \qquad (8)$$

$$h^{act} = c^i \Omega (B - T\beta B - T\frac{dB}{dT}\bigg|_P) \qquad (9)$$

where $\beta$ is the thermal (volume) expansion coefficient.

## 3. Application to the case of W diffusing in $\alpha$-Fe

We now apply Eq.(7) to the case of tungsten diffusing in $\alpha$-Fe. Concerning the attempt frequency, $\nu$, we consider that for a given matrix and mechanism, it depends roughly on the mass of the diffusant according to the approximation:

$$\frac{\nu^j}{\nu_D} = \left(\frac{m^m}{m^j}\right)^{\frac{1}{2}} \qquad (10)$$

where $m^m$, $m^j$ denote the mass of the matrix (*m*) and the diffusant (*j*), respectively (i.e., Fe and W in the present case) and $\nu_D \approx 9 \times 10^{12} \text{s}^{-1}$. Concerning the elastic data, the adiabatic bulk modulus has been measured in the region 298 to 1173 K by Dever [18] and is converted to the isothermal one, *B*, by means of the expansivity and specific heat data given in the literature (see Ref. [17] and references therein). The determination of $c^{act}$ is now made at the temperature $T_I$=973 K for which Takemoto et al. [8] reported two measurements for *D*, i.e., $1.91 \times 10^{-18}$ and $1.81 \times 10^{-18}$ m²/s. Hence, we use here their average value, i.e., $D_I = 1.86 \times 10^{-18}$ m²/s and also consider that – according to the elastic data mentioned above- *B*=133.3 GPa at this temperature; furthermore, we take into account that $\alpha \approx 2.89$Å (cf. recall that $\Omega = \alpha^3/2$) and assume that the diffusion proceeds via monovacancies, thus *f*=0.727. By inserting these values into Eq.(7), we find that $c^{act}$ has a value between 0.21 and 0.22 after considering plausible experimental errors in the quantities used in the calculation.

Once $c^{act}$ is known, we can now compute *D* for every temperature by incorporating the appropriate data of *B* and $\Omega$ into Eq.(7). The calculation was made at all temperatures (between 833 K and $T_C$) at which experimental *D* values have been reported by Takemoto et al. [8] by using the *B* and $\Omega$ values resulting from a linear interpolation of the corresponding experimental values given in Table 1 of Ref. [17]. These calculated *D* values are inserted with stars in Fig. 1, while the experimental values are shown with open circles. Note that, since $c^{act}$ was taken as 0.21 or 0.22, as mentioned above, two calculated $D$ values are depicted for each temperature. An inspection of this figure reveals $\Omega$ that the experimental $D$ values lie more or less between the calculated ones.

We now calculate $h^{act}$, for example at the temperature $T$=993 K, in which $B$=132.2 GPa, $\beta\approx 5\times 10^{-5}$K$^{-1}$ and $\Omega\approx 12.13\times 10^{-30}$m$^3$. Furthermore, we consider that (dB/dT)=-0.0624 GPa/K as it results from a least squares fitting to a straight line of the $B$-values given in Table 1 of Ref. [17] between 973 and 1043 K. Inserting these values into Eq.(9), we find $h^{act}$=3.0 and 3.13 eV for $c^{act}$=0.21 and 0.22 respectively, which are in excellent agreement with the experimental value [8] $h^{act}$=(3.0±0.2) eV.

## 4. Discussion

We now discuss the following empirical fact mentioned in Ref. [8]. Studying the diffusion of transition elements, such as Ti, V, Cr, Co, Ni, Nb, Mo and W in paramagnetic $\alpha$-iron, the activation enthalpy $h^{act,j}$ was found to increase linearly with $(r_{solute}-r_{Fe})/r_{Fe}$, where $r_{solute}$ and $r_{Fe}$ are the radii of the solute atom for the coordination number eight and of iron matrix lattice. In other words, atomic size affects the activation enthalpies for diffusion of transition elements in paramagnetic $\alpha$-iron. This is strikingly reminiscent of an early finding in alkali halides doped with divalent cations, in which electric dipoles of the form "divalent cation plus one cation vacancy" are produced [19]. These dipoles, upon applying an external electric field, change their orientation in space mainly through jumps of the cation vacancy between neighboring sites to the divalent impurity (cf. these dipoles contribute of course to the static dielectric constants $\varepsilon_S$, but even in their absence (i.e., in the case of "pure" alkali halides) $\varepsilon_S$ varies upon changing the temperature or pressure, mainly due to the volume dependence of the ionic polarizability which is interrelated with $B$ [20]). The activation enthalpy for this (re)orientation process, which is of course governed by the vacancy migration, was found [21] to increase linearly with the ionic radius of the divalent cations, when the latter have rare gas electron configuration.

The $cB\Omega$ model cannot give any direct explanation for the aforementioned effect, either in paramagnetic $\alpha$-iron or in alkali halides. Only an indirect guess for the effect in $\alpha$-iron could be made along the following lines: First, let us make the reasonable assumption that a diffusant having larger atomic size corresponds to a larger activation volume $\upsilon^{act,\,j}$. Second, by inserting Eq.(6) into the relation [1] $\upsilon^{act} = \left.\dfrac{dg^{act}}{dP}\right|_T$, we find:

$$\upsilon^{act,j} = c^{act,j}\left(\left.\frac{dB}{dP}\right|_T - 1\right)\Omega \qquad (11)$$

which, when combined with Eq.(9), leads to the conclusion that the ratio $\upsilon^{act,\,j}/h^{act,\,j}$ is a bulk quantity, i.e.,

$$\frac{\upsilon^{act}}{h^{act}} = \frac{\left.\dfrac{dB}{dP}\right|_T - 1}{B - T\beta B - T\left.\dfrac{dB}{dT}\right|_P} \qquad (12)$$

and hence should be the same for various diffusants $j$ in the same matrix. Thus, on the basis of $cB\Omega$ model, we can guess that a diffusant with larger atomic size should also have a larger activation enthalpy. A quantitative assessment in terms of the atomic radius cannot be made.

## 5. Conclusion

The diffusion coefficient of W in $\alpha$-iron can be satisfactorily calculated in the temperature range from 833 K to $T_C$(=1043 K) upon employing the $cB\Omega$ model. This calculation is made without using any adjustable parameter.

**FIGURE and FIGURE CAPTION**

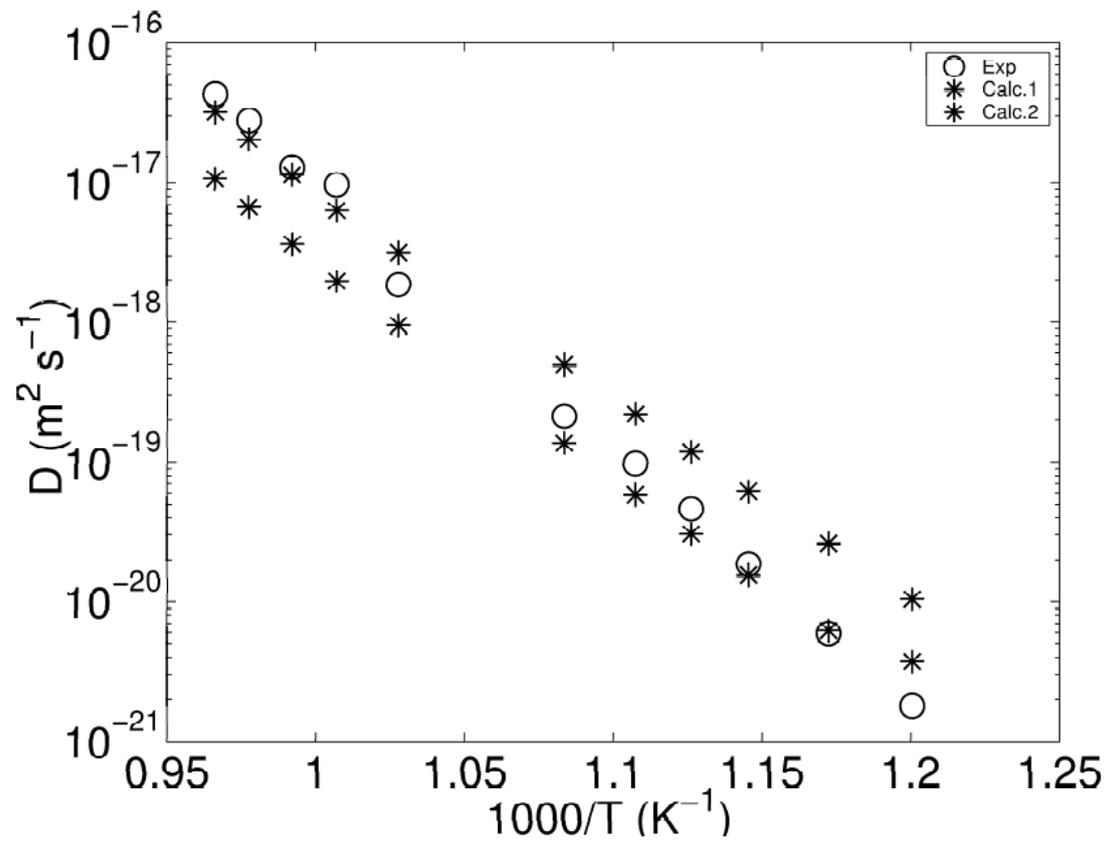

**Fig. 1.** Tungsten diffusing in $\alpha$-iron. Diffusion coefficients, *D*, as measured in Ref. [8] (circles) at various temperatures, T, vs 1000/T. The upper and the lower value calculated for each temperature by means of the $cB\Omega$ model are shown by stars.